\documentclass[5p,twocolumn]{elsarticle}

\usepackage[utf8]{inputenc}
\usepackage{graphicx}
\usepackage{verbatim}
\usepackage[T1]{fontenc}
\usepackage[english]{babel}
\usepackage{bm}
\usepackage{enumerate}
\usepackage{amsmath}
\usepackage{amssymb}
\usepackage{amsbsy}
\usepackage{amsthm}
\usepackage{bbold}
\usepackage{soul}
\usepackage{color}
\usepackage{subfigure}
\usepackage[shortcuts]{extdash}
\usepackage{upgreek}
\usepackage{mathtools}
\usepackage[colorlinks=true, citecolor=blue]{hyperref}
\usepackage[normalem]{ulem}
\usepackage[dvipsnames]{xcolor}

\DeclareMathOperator{\Tr}{Tr}
\newcommand{\qhat}{\hat{q}}

\newcommand{\dg}{\dagger}
\newcommand{\ev}[1]{\langle  #1 \rangle }

\newcommand{\p}{\partial}

\newcommand{\A}{\mathrm{A}}
\newcommand{\B}{\mathrm{B}}

\begin{document}
\title{Jet momentum broadening in the pre-equilibrium Glasma}

\author[1]{A.~Ipp}
\ead{ipp@hep.itp.tuwien.ac.at}
\author[1]{D.~I.~Müller\corref{cor1}}
\ead{dmueller@hep.itp.tuwien.ac.at}
\cortext[cor1]{Corresponding author}
\author[1]{D.~Schuh}
\ead{schuh@hep.itp.tuwien.ac.at}
\address[1]{Institute for Theoretical Physics, TU Wien, \\
Wiedner Hauptstr. 8-10, A-1040 Vienna, Austria}
\date{\today}

\begin{abstract}
    Jets are important probes of heavy ion collisions as they can provide information on the interaction of a
    highly energetic parton with the medium it traverses. 
    In the hydrodynamic stage of dense, strongly interacting matter, these interactions can be explained in terms of scattering processes,
    soft gluon emission and collinear parton splittings.
    However, jets originate even before the hydrodynamic stage.
    Here we report on the first numerical simulation of transverse momentum broadening of jets stemming from the interaction
    of partons with boost-invariantly expanding Glasma flux tubes.
    The Glasma stage is a pre-hydrodynamic stage based on the Color Glass
    Condensate framework.
    Our calculation shows strong time-dependence and an intrinsic anisotropy of momentum broadening in the directions transverse to the jet propagation direction.
\end{abstract}


\maketitle

\section{Introduction}

Heavy-ion collisions at the Large Hadron Collider (LHC) or at the Relativistic Heavy Ion Collider (RHIC) provide a unique opportunity to study quantum chromodynamics (QCD)
under extreme conditions.
The quark-gluon plasma (QGP) that is created in such collisions
can be well described by relativistic viscous hydrodynamics
\mbox{\cite{Gale:2013da, Romatschke:2017ejr}},
which accurately predicts particle multiplicities and flow harmonics \mbox{\cite{Schenke:2010nt, Schenke:2011bn, Schenke:2012fw, Gale:2012rq, Niemi:2015qia}}. 
Due to the existence of a hydrodynamic attractor \cite{Berges:2013fga}, such bulk properties are rather 
insensitive to the details of the initial conditions of the hydrodynamical evolution.
There are however probes that carry information throughout the whole
history of the evolution. These include probes which do not interact strongly, like photons and dileptons 
\mbox{\cite{Mauricio:2007vz, Ipp:2009ja, Ipp:2012zb, Vujanovic:2014xva}},
but also strongly interacting hard probes with momenta much larger than typical momenta of the medium,
known as jets \mbox{\cite{Mehtar-Tani:2013pia, Connors:2017ptx, Busza:2018rrf}}. 
Since jets are created in the initial stage of heavy-ion collisions, they are affected by the entire space-time evolution of the medium, including its pre-equilibrium stage.

Recently, the pre-equilibrium stage and its effects on jet observables garnered increasing attention. 
In \cite{Andres:2019eus}, simulations of relativistic hydrodynamics with initial conditions provided by the EKRT framework \cite{Eskola:1999fc} 
suggest that high $p_T$ harmonics can be sensitive to the details of the jet evolution in the initial stage.
A systematic treatment of the pre-equilibrium stage is possible within the Color Glass Condensate (CGC) framework \mbox{\cite{Gelis:2010nm, Gelis:2012ri}}. In this framework, non-Abelian interactions lead to the formation of a highly anisotropic state of expanding chromo-electric and chromo-magnetic flux tubes known as the Glasma \cite{Lappi:2006fp}.
The possible effect of this stage has been estimated using various methods \mbox{\cite{Schenke:2008gg, Aurenche:2012qk, Sun:2019fud, Liu:2019lac}}, but only recently a more systematic small proper time approximation was performed to calculate energy loss and momentum broadening from the Glasma \cite{Carrington:2020sww}. However, to go beyond very small proper times requires the use of lattice methods which fully respect gauge invariance.

In this Letter we present our non-perturbative SU(3) real-time lattice simulation results for the transverse momentum broadening contribution to a high energy parton
from the boost-invariant Glasma phase. Our approach is valid for larger proper times and
gauge invariant. We show that the Glasma gives a sizeable contribution to momentum broadening and that it occurs anisotropically with larger broadening along rapidity.
Details and various checks, including the comparison of lattice results with analytical calculations in the dilute limit can be found in \cite{Ipp:2020b}.

\section{Theoretical framework}
The theoretical basis for our calculation is given by the CGC framework, which we review and summarize in this section. 
For more details regarding the CGC and the Glasma, we refer to \mbox{\cite{Lappi:2006fp, Gelis:2010nm, Gelis:2012ri}}.
Within the CGC framework, partons of the incoming nuclei with high momentum fraction
$x$ are described as static, classical color charges. The color current of a nucleus ``A'' (``B'') moving along the positive (negative) beam axis is given by
\begin{equation}
J^\mu_{(\A,\B)} = \delta^\mu_\pm \rho_{(\A,\B)}(x^\mp, \mathbf x),
\end{equation}
where $x^\pm = (t \pm z) / \sqrt{2}$ and $\mathbf x = (x, y)$ is the transverse coordinate vector. The large Bjorken~$x$ partons act as sources for the low
$x$ partons, which are described by a classical color field $A^\mu$. The color field is determined by the Yang-Mills (YM) equations
\begin{equation}
D_\mu F^{\mu\nu}_{(\A,\B)} = J^\nu_{(\A,\B)}. \label{eq:YM}
\end{equation}
In light cone gauge $A^+ = 0$ ($A^- = 0$), the color field is purely transverse
\begin{equation}
A^i_{(\A,\B)}(x^\mp, \mathbf x) = \frac{1}{ig} V_{(\A,\B)}(\mathbf x) \p^i V_{(\A,\B)}(\mathbf x) \theta(x^\mp),
\end{equation}
with Wilson lines $V$ given by
\begin{equation}
V^\dg_{(\A,\B)}(\mathbf x) = \mathcal{P} \exp{\left(i g \intop^{\infty}_{-\infty} dx^\mp \frac{\rho_{(\A,\B)} (x^\mp , \mathbf x)}{\mathbf \nabla^2 - m^2}\right)}. \label{eq:CGC_Wilson}
\end{equation}
Here, $m$ acts as an infrared regulator usually taken to be roughly $\Lambda_\mathrm{QCD}$ and $g$ is the YM coupling constant. For ${m=0}$ the infrared divergence is regulated by requiring color neutrality of the whole nucleus \cite{KRASNITZ2003268}. 
The initial Glasma field created at proper time ${\tau = \sqrt{2 x^- x^+} = 0}$ is given in temporal gauge by \cite{Kovner:1995ja}
\begin{align}
A^i(\mathbf x) &= A^i_\A(\mathbf x) + A^i_\B(\mathbf x), \label{eq:g_ic1}\\
A^\eta(\mathbf x) &= \frac{ig}{2} \left[ A^i_\A(\mathbf x), A^i_\B(\mathbf x)\right]. \label{eq:g_ic2}
\end{align}
For $\tau > 0$, the Glasma evolves in time according to the source-free YM equations 
\begin{equation}
    D_\mu F^{\mu\nu} = 0. \label{eq:YM_sf}
\end{equation}

\begin{figure}
    \centering
    \includegraphics[scale=0.2]{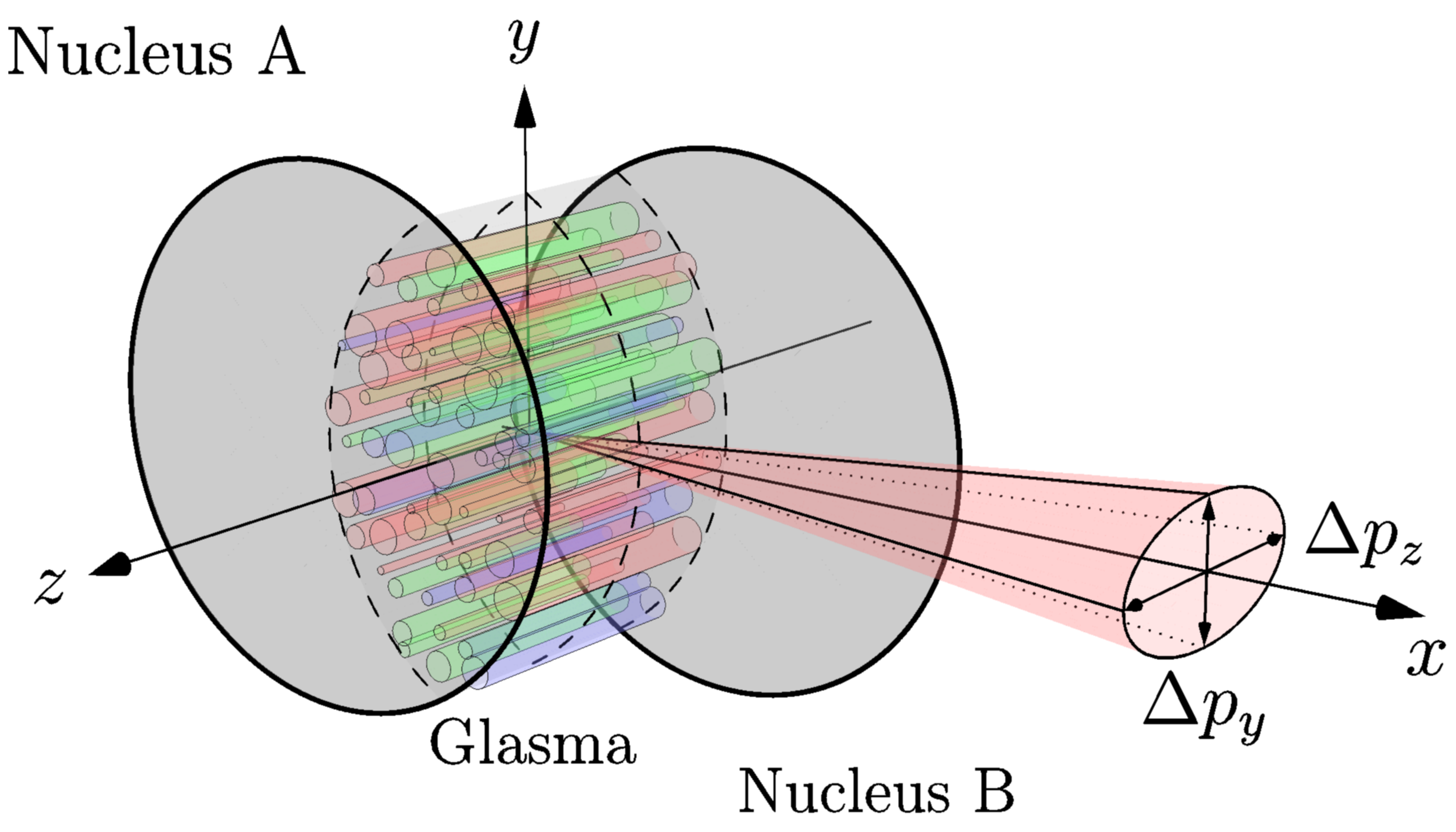}
    \caption{
    Geometry of a heavy-ion collision including an emitted particle jet (from \cite{Ipp:2020b}).
    Depending on the context, ``transverse'' and ``longitudinal'' can
    refer to different axes:
    regarding the whole collision event, the transverse directions are those
    orthogonal to the beam axis ($z$-axis), along which the nuclei collide. 
    When referring to momentum broadening, transverse momenta
    are orthogonal to the jet propagation direction ($x$-axis).
    Anisotropic momentum broadening thus compares momentum broadening
    along the $y$- and $z$-axes, 
    which corresponds to broadening along the azimuthal ($\Delta \phi$) and rapidity ($\Delta\eta$) directions.
    }
    \label{fig:momentum_broadening}
\end{figure}

The color charge density $\rho$ of a nucleus is treated as a random field, distributed according to a probability functional $W[\rho]$. Expectation values of observables are computed by averaging over all configurations weighted by $W[\rho]$. In this work we apply the McLerran-Venugopalan (MV) model \mbox{\cite{MV1, MV2}}, where $W[\rho]$ is Gaussian and determined by the charge density correlator 
\begin{equation}
\ev{\rho^a_{\mspace{-2mu}(\mspace{-2mu} \A, \B \mspace{-2mu})} \mspace{-2mu} (\mspace{-1mu} x \mspace{-1mu}) \rho^b_{\mspace{-2mu}(\mspace{-2mu} \A, \B \mspace{-2mu})}(\mspace{-1mu} y \mspace{-1mu}) \mspace{-2mu}} \mspace{-1mu} = \mspace{-1mu} (g\mu \mspace{-1mu} )^{\mspace{-2mu} 2} \mspace{-2mu} \delta^{ab} \mspace{-2mu} \delta( \mspace{-1mu} x^{\mp} \mspace{-7mu} - \mspace{-1mu} y^{\mp}) \delta( \mspace{-1mu} x^{\mp}) \delta^{(2)} \mspace{-3mu} ( \mspace{-1mu} \mathbf x \mspace{-1mu} - \mspace{-1mu} \mathbf y \mspace{-2mu} ).
\end{equation}
The MV model parameter $\mu$ fixes the saturation momentum $Q_s \propto g^2 \mu$. 
Additionally, the color neutrality of nuclei is guaranteed by requiring $\ev{\rho^a(x)} = 0$.

\section{Momentum broadening in the Glasma}

The color fields of the Glasma can exert strong Lorentz forces on highly energetic partons that originate from initial hard scatterings and eventually become jets.
Here we review the main steps to calculate transverse momentum broadening from the Glasma.
We consider a single quark emitted at the origin of the collision $x^+=x^-=0$ with very high initial momentum along the $x$ axis and vanishing momentum rapidity along $z$ (see fig.~\ref{fig:momentum_broadening}).
The jet broadening parameter $\qhat_\perp$ is generally defined as the accumulated squared transverse momentum per unit time (or equivalently per unit length).
We define the instantaneous jet broadening parameter $\qhat_i(\tau)$ for the transverse direction $i \in \{ y, z \}$ as
\begin{equation}
\qhat_i(\tau) = \frac{d}{d\tau} \ev{p^2_i(\tau)}_q.
\end{equation}
The momentum change of the quark is due to interactions with the classical background Glasma field which can be modeled using Wong's equations
\begin{align}
\frac{dp_\mu}{d\tau} &= g Q^a(\tau) \frac{d x^\nu}{d\tau} F^{a}_{\mu\nu}(\tau), \label{eq:wong_1}\\
\frac{d Q^a}{d\tau} &= g \frac{dx^\mu}{d\tau} f^{abc} A^b_\mu(\tau) Q^c(\tau), \label{eq:wong_2}
\end{align}
where $x^\mu(\tau)$ is the trajectory and $Q^a(\tau)$ is the color charge of the quark.
The fields are understood to be evaluated along the particle trajectory, which we assume to be lightlike for a highly energetic quark. The quark is considered to be a test particle, i.e.~it is affected by the color fields of the Glasma, but the Glasma itself is unaffected by the quark traversing it. Since the background field is invariant under boosts and the test quark is ultrarelativistic, the momentum broadening does not change even if a different, non-zero initial rapidity of the quark were assumed.
Integrating eqs.~\eqref{eq:wong_1} and \eqref{eq:wong_2}, we find the following gauge invariant expression
\begin{equation}
\ev{p^2_i(\tau)}_q = \frac{g^2}{N_c} \intop^\tau_0 d\tau' \intop^\tau_0 d\tau'' \ev{\mathrm{Tr} \left[ f^i(\tau') f^i(\tau'') \right]}, \label{eq:acc_mom}
\end{equation}
where no sum over $i$ is implied on the right-hand side. Here, $f^i(\tau')$ is the color-rotated Lorentz force given by
\begin{align}
f^y(\tau) &= U(\tau) \left( E_y(\tau) - B_z(\tau) \right) U^\dg(\tau), \label{eq:fy_EB}\\
f^z(\tau) &= U(\tau) \left( E_z(\tau) + B_y(\tau) \right) U^\dg(\tau), \label{eq:fz_EB}
\end{align}
where $E$ and $B$ are the color-electric and \-/magnetic fields evaluated at the particle position $x^\mu(\tau)$. The matrix $U(\tau)$ is a lightlike Wilson line in the fundamental representation along the particle trajectory. It accounts for the color rotation of the quark as it moves through the color field of the Glasma. In temporal gauge $A^\tau = 0$, it is given by
\begin{equation}
U(\tau, 0) = \mathcal{P} \exp{\bigg( - i g \intop_0^\tau d\tau' A_x (\tau') \bigg)}. \label{eq:ll_U}
\end{equation}

Alternatively, eq.~\eqref{eq:acc_mom} can be derived using the dipole approximation \cite{Liu:2006ug}. The expectation value of a particular rectangular Wilson loop $W_i$ with lengths $L_\perp \ll L^+$ is related to the transverse momentum $\ev{p^2_i(\tau)}_q$ of a quark:
\begin{equation}
\frac{1}{N_c} \ev{\operatorname{Re} \Tr \left[ W_{i} \right] } \approx \exp{\bigg( -\frac{L_\perp^2}{2} \ev{p^2_i(\tau)}_q \bigg)}. \label{eq:wilson}
\end{equation}
This Wilson loop consists of four straight Wilson lines: two lightlike Wilson lines of length $L^+$ (following the particle trajectory as in eq.~\eqref{eq:ll_U}), separated at a distance $L_\perp$ along one of the jet-orthogonal directions $i \in  \{ y, z \}$, and two spatial Wilson lines closing the loop. 
In \cite{Ipp:2020b} we show that a Taylor expansion in $L_\perp$ yields the same result as integrating the classical equations of motion for a colored particle, namely eq.~\eqref{eq:acc_mom}.

\begin{figure}[t]
    \centering
    \includegraphics{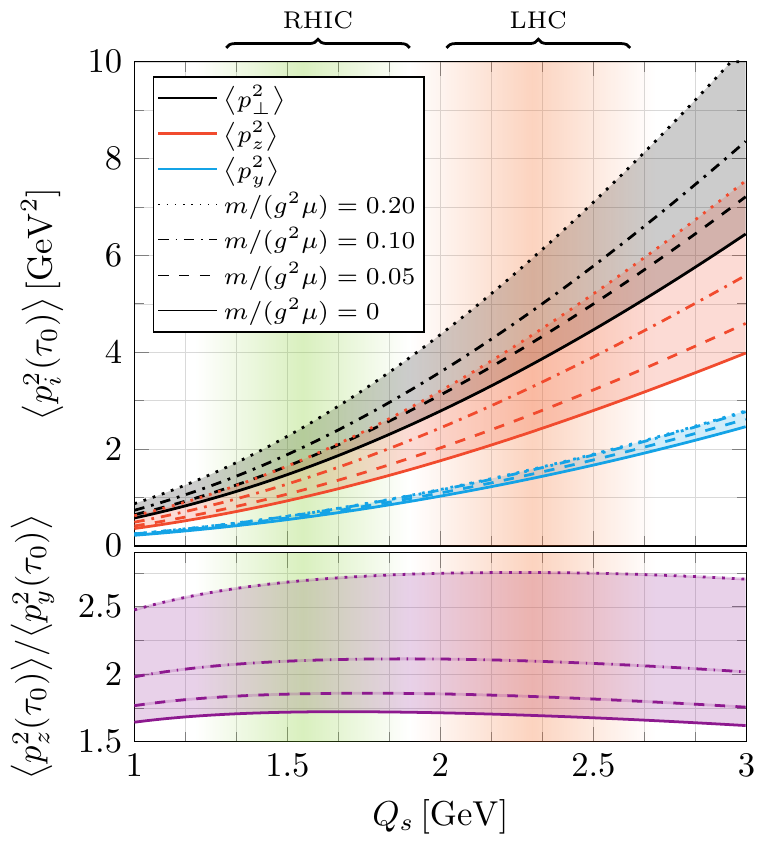}
    \caption{Accumulated transverse momenta $\ev{p^2_i}$ (top plot) and the momentum broadening anisotropy $\ev{p^2_z} / \ev{p^2_y}$ (bottom plot) at $\tau_0 = 0.6 \, \mathrm{fm}/c$ as a function of saturation momentum $Q_s$ for a high energy quark moving along the $x$ axis (see fig.~\ref{fig:momentum_broadening}). The upper plot contains $\ev{p^2_y}$ (blue, lower band), which corresponds to $\Delta \phi$ broadening, $\ev{p^2_z}$ (orange, middle band), corresponding to $\Delta \eta$ broadening, and the total transverse momentum $\ev{p^2_\perp} = \ev{p^2_y} + \ev{p^2_z}$ (black, upper band). The uncertainty in the results corresponds to different values of the infrared regulator $m$. The colorful vertical bands indicate which values of $Q_s$ are relevant to RHIC and LHC physics.}
    \label{fig:acc_mom_Qs}
\end{figure}

Generally, eq.~\eqref{eq:acc_mom} is hard to evaluate analytically: the non-Abelian background field $A_\mu$ enters in a highly non-linear way through the Wilson lines and $A_\mu$ itself has to be determined from the YM eqs.~\eqref{eq:YM_sf}. To proceed, one has to either treat $A_\mu$ perturbatively \cite{Kovner:1995ts}, which corresponds to the case of a dilute Glasma, or evaluate eq.~\eqref{eq:acc_mom} numerically using lattice simulations \mbox{\cite{Krasnitz:1998ns, Lappi:2003bi}}.  
In \cite{Ipp:2020b}, we have performed both a weak field expansion for the dilute Glasma and formulated the problem in terms of real-time lattice gauge theory to investigate transverse momentum broadening in the dense Glasma.

These results can be easily extended to gluon jets due to Casimir scaling. The accumulated squared momenta and the jet broadening parameter for gluons are given by
\begin{align}
\ev{p^2_\perp}_g = \frac{C_A}{C_F} \ev{p^2_\perp}_q, \qquad \qhat_g(\tau) = \frac{C_A}{C_F} \ \qhat_q(\tau),
\end{align}
where $C_A$ and $C_F$ are the Casimirs in the adjoint and fundamental representation respectively. For $N_c=3$ one finds $C_A / C_F = 9 / 4$. 

\section{Results and discussion}
In this section we present our results of transverse momentum broadening of quarks from SU(3) real-time lattice simulations of the Glasma.
In fig.~\ref{fig:acc_mom_Qs} (top) we plot the accumulated squared transverse momenta $\ev{p^2_i}$ evaluated at proper time $\tau_0 = 0.6\,\mathrm{fm}/c$ as a function of $Q_s$ for various values of the infrared regulator $m$.
This value of $\tau_0$ corresponds to the typical starting time of jet energy loss calculations which neglect pre-equilibrium effects \cite{Andres:2019eus}.
Due to the simplicity of the MV model the only relevant parameter is the ratio $m / (g^2 \mu)$ with $m / (g^2 \mu) \ll 1$ corresponding to the dense Glasma produced in relativistic heavy-ion collisions and $m / (g^2 \mu) \gg 1$ corresponding to the dilute Glasma.
For $m = 0$ the infrared divergence of the MV model is regulated by the system size given by $g^2 \mu L =  100$, where $L$ is the transverse extent of our simulation box. We use a transverse lattice of size $1024^2$ and $N_s = 50$ color sheets for the longitudinal discretization of the MV model \cite{Fukushima:2007ki}.
Using the numerical results of \cite{Lappi:2007ku} the values of $g^2\mu$ can be converted to saturation momenta $Q_s$. 
We find that the total squared transverse momentum $\ev{p^2_\perp} = \ev{p^2_y} + \ev{p^2_z}$ picked up by a high energy quark is roughly $\ev{p^2_\perp} \approx Q^2_s$, although our results depend on the exact value of $m$ (shown as shaded regions in fig.~\ref{fig:acc_mom_Qs}). Interestingly, the momentum broadening of jets in the Glasma happens anisotropically: a high energy quark receives more momentum along the beam axis (rapidity broadening) compared to broadening in the transverse plane of the Glasma (azimuthal broadening). We show the momentum broadening anisotropy $\ev{p^2_z} / \ev{p^2_y}$ as a function of $Q_s$ in fig.~\ref{fig:acc_mom_Qs} (bottom). The anisotropy is largely independent of $Q_s$ but shows strong dependence on $m / (g^2 \mu)$ with higher anisotropy for more dilute Glasmas. 

\begin{figure}
    \centering
    \includegraphics[scale=0.25]{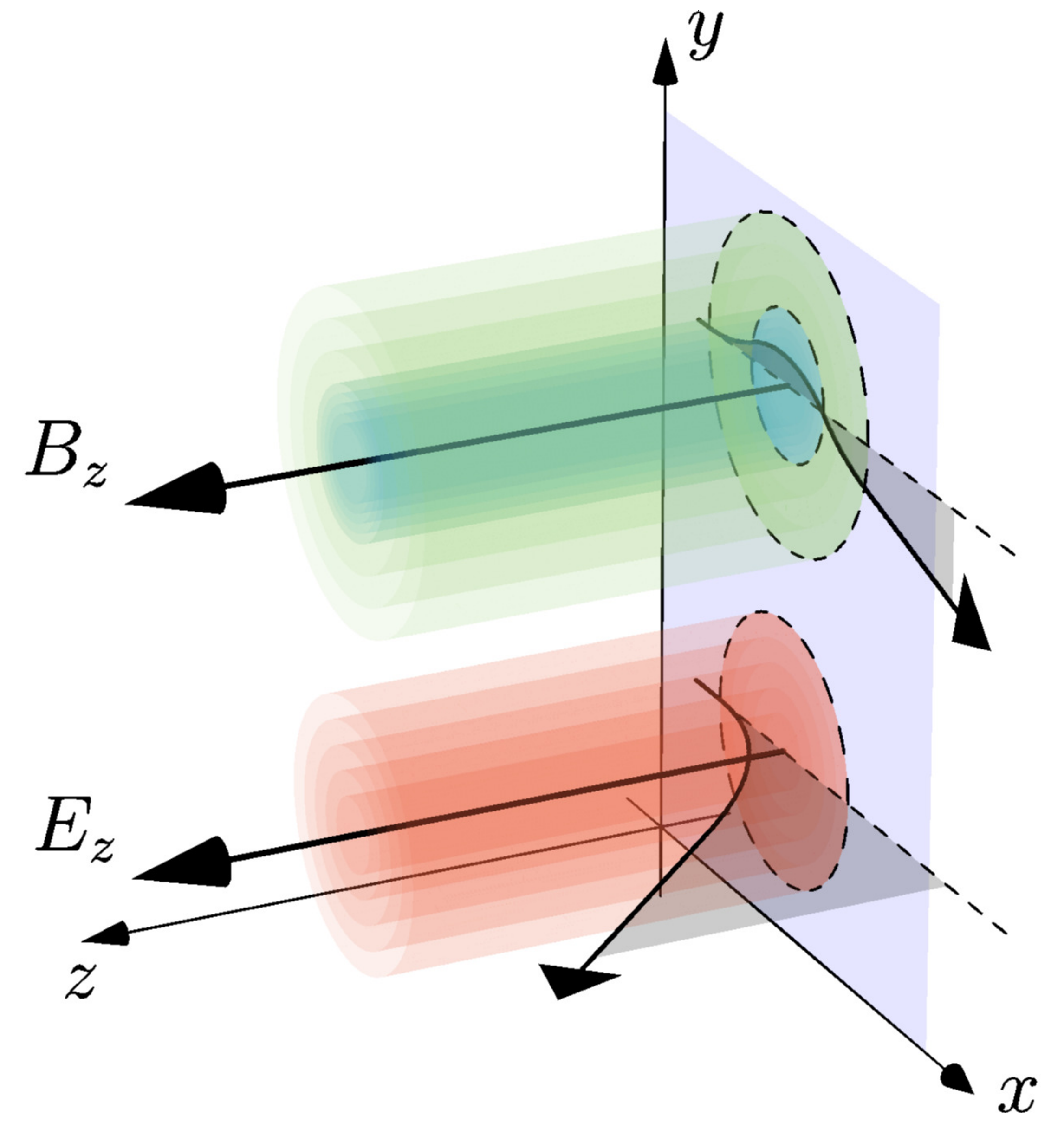}
    \caption{
        Physical origin of the momentum broadening in the Glasma. A high-energy quark is differently accelerated in a color-electric Glasma flux tube (red) and a color-magnetic Glasma flux tube (blue and green). Although electric and magnetic flux tubes are of similar size and field strength, electric flux tubes consist of uniformly oriented fields, while magnetic flux tubes exhibit a ring of anti-correlated fields with opposite sign (green ring around blue center). This leads to a suppression of momentum broadening along the $y$-axis (azimuthal direction) compared to the $z$-axis (rapidity direction).
    }
    \label{fig:flux_tubes}
\end{figure}

To investigate the origin of the anisotropy, it is useful to look at the case of a dilute Glasma, where a perturbative calculation is possible. 
In this limit, eq.~\eqref{eq:acc_mom} simplifies to \cite{Ipp:2020b}
\begin{equation}
\ev{p^2_{(y, z)}(\tau)} = \intop^\infty_0 dk \, g(\tau, k) \, c_{(E, B)}(k), \label{eq:dilute_acc_mom}
\end{equation}
where $g(\tau, k)$ is a function that describes the time evolution of the Glasma and $c_{(E, B)}(k)$ are the Fourier components of the longitudinal electric or longitudinal magnetic correlator at $\tau = 0$
\begin{align}
c_E(r) &= \ev{ \mathrm{Tr} \left[ E_z(\mathbf x) E_z(\mathbf y) \right]}, \label{eq:Ecorr} \\
c_B(r) &= \ev{ \mathrm{Tr} \left[ B_z(\mathbf x) B_z(\mathbf y) \right]}, \label{eq:Bcorr}
\end{align}
with $r = |\mathbf x - \mathbf y|$. 

Since the only directional dependence in eq.~\eqref{eq:dilute_acc_mom} enters through these correlators, this yields a remarkably simple picture of jet momentum broadening in terms of Glasma flux tubes: the initial Glasma directly after the collision is composed of color-electric and \-/magnetic flux tubes aligned with the beam axis $z$. The radius $r_s$ of a typical flux tube is roughly $Q_s^{-1}$ and both electric and magnetic flux tubes contribute roughly equally to the initial energy density of the Glasma \cite{Lappi:2006fp}.
A high energy quark moving through this initial Glasma along the $x$ direction receives additional momentum along $z$ from electric flux tubes and momentum along $y$ from magnetic flux tubes according to the non-Abelian Lorentz force.
Interestingly, even as these two kinds of flux tubes continue to evolve and expand, their bending effect stays the same. Let us consider a part of the expanding wave that propagates along the $x$ direction. The originally longitudinal electric flux tube along $z$ turns into a polarized wave mode with electric fields aligned along $z$ and magnetic fields aligned along $y$, but both fields lead to a bending of the hard particle along $z$.
Similarly, the originally longitudinal magnetic flux tube leads to a polarization mode that always bends a particle along the $y$ direction.
A hard particle that originates close to a flux tube will co-propagate with its expanding wave and will thus have a long time to acquire the momentum change in a particular direction as dictated by the initial type, orientation and strength of flux tube.

The observation of a momentum broadening anisotropy in the Glasma can be therefore traced back to a difference in the initial shapes of electric and magnetic Glasma flux tubes as seen in fig.~\ref{fig:flux_tubes}. The shapes of Glasma flux tubes are characterized by the respective correlation functions eqs.~\eqref{eq:Ecorr} and \eqref{eq:Bcorr}. We found that magnetic flux tubes show a pronounced region of anti-correlation at roughly $r \approx 1.5 \, Q^{-1}_s$, where the orientation of the magnetic field $B_z$ is flipped compared to the center of the flux tube (this has also been observed in \cite{Dumitru:2014nka} and \cite{Ruggieri:2017ioa}). In contrast, electric flux tubes show no such anti-correlation or a much less pronounced, weaker region around $r \approx 2 Q^{-1}_s$. As a result, shown in fig.~\ref{fig:flux_tubes}, a quark moving through such Glasma flux tubes can more efficiently accumulate momentum along $z$ from electric flux tubes, leading to the observed momentum broadening anisotropy in fig.~\ref{fig:acc_mom_Qs}. The effect of the anisotropy is further enhanced by the simultaneous expansion of the flux tubes in the $xy$-plane \cite{Ipp:2020b}.

Early anisotropic momentum broadening is a possible explanation for observable effects like a ridge-like structure in dihadron correlations \cite{Mizukawa:2008tq}, but the emergence of momentum anisotropy is not restricted to the Glasma. 
Anisotropic broadening in the QGP phase has been investigated using kinetic theory \cite{Romatschke:2006bb}, including the effect of plasma instabilities \mbox{\cite{Majumder:2006wi,Dumitru:2007rp}}, and the gauge/gravity correspondence \mbox{\cite{2012JHEP...07..031G,2012JHEP...08..041C, Rebhan:2012bw}}.
In all these scenarios, jets receive stronger broadening along the beam axis, as we do, but
the cause of anisotropy is an anisotropy in the momentum distribution of the QGP.
In contrast, the physical origin of anisotropic broadening from the Glasma is due to anti-correlated regions around color-magnetic Glasma flux tubes. 

\begin{figure}
    \centering
    \includegraphics{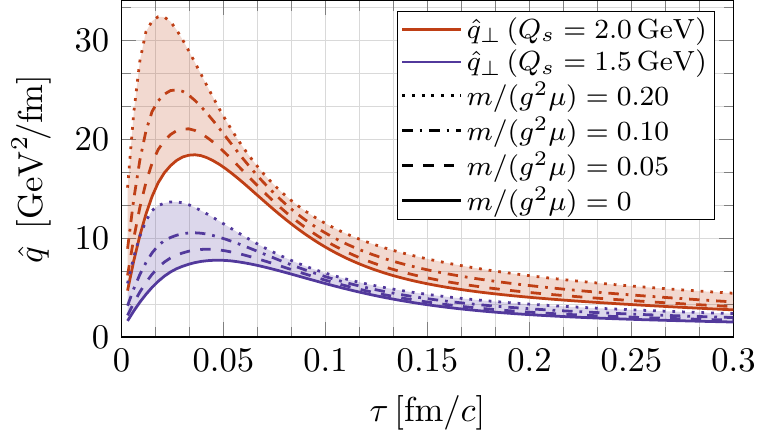}
    \caption{Jet broadening parameter $\qhat_\perp$ for quarks as a function of proper time $\tau$ for two different values of the saturation momentum $Q_s$. The red lines correspond to $Q_s=2 \, \mathrm{GeV}$ and the blue lines to $Q_s=1.5 \, \mathrm{GeV}$. We see that most momentum is acquired in the very early stage $\tau < 0.1 \, \mathrm{fm}/c$. After this initial strong broadening, the jet quenching parameter drops very quickly to lower values.  }
    \label{fig:qhat_tau}
\end{figure}

Finally, we also present our numerical results for the jet broadening parameter $\qhat_\perp$ as a function of proper time $\tau$ in fig.~\ref{fig:qhat_tau}. 
We observe a strong time dependence when we plot the instantaneous broadening parameter $\qhat_\perp$ for two typical values of the saturation momentum $Q_s \in \{ 1.5 \, \mathrm{GeV}, 2.0 \, \mathrm{GeV} \}$. As can be seen from the plot, a fast quark receives a large amount of transverse momentum in the first $\tau \approx 0.1 \, \mathrm{fm} / c$ which corresponds to $\tau \approx Q^{-1}_s$. After this initial broadening, the jet broadening parameter $\qhat_\perp$ quickly drops and the accumulation of transverse momentum is slowed down. Therefore, even though $\qhat_\perp$ peaks at very high values, a quark only obtains $3 - 4 \, \mathrm{GeV^2}$ of squared transverse momentum until $\tau = 0.6 \, \mathrm{fm} / c$. At these later times, the classical field approximation for the Glasma is less accurate and we expect other effects of later stages to become more important.

The results shown in fig.~\ref{fig:qhat_tau} can be extended to other values of $Q_s$: since $Q_s$ is the only relevant dimensionful scale in the dense Glasma (for a fixed ratio $ m / g^2 \mu \lesssim 1$), $\qhat_\perp$ can be rescaled to other values of $Q'_s$ by performing $\qhat_\perp \rightarrow (Q'_s / Q_s)^3 \qhat_\perp$ and $\tau \rightarrow (Q'_s / Q_s) \tau$.

\section{Conclusions and Outlook}

In this Letter we have presented our results on transverse momentum broadening of high energy quarks traversing the pre-equilibrium Glasma. We have performed non-perturbative real-time lattice simulations which are by construction gauge invariant. Our main findings are that quarks accumulate up to $\ev{p^2_\perp} \approx Q^2_s$ of transverse momentum within a short time span of $< 0.6 \, \mathrm{fm}/c$. The broadening of quarks is anisotropic with more efficient broadening along rapidity, compared to the azimuthal directions. The origin of this anisotropy lies in the spatial correlation structure of the initial Glasma flux tubes. 

We have also computed the jet broadening parameter $\qhat_\perp$ and shown that it strongly depends on time with the largest contributions  stemming from the first $0.1 \, \mathrm{fm}/c$.

One of the effects that has not been accounted for in this work is that of parton energy loss, i.e.~the loss of kinetic energy as quarks or hard gluons pass through the Glasma, which we plan to investigate in the future. It would also be interesting to better understand the physics behind the initial correlations of color-electric and \-/magnetic Glasma flux tubes, which lead to the momentum broadening anisotropy. 
A straightforward and relevant extension of our work would be to study momentum broadening in a non-boost-invariant setup \mbox{\cite{Schenke:2016ksl,Gelfand:2016yho,Ipp:2017lho,Ipp:2018hai,McDonald:2018wql, Ipp:2020igo}}.
Most importantly however, our results should be used as input for simulations of the later stages of jet evolution to study the full phenomenological implications of the Glasma stage on jet observables.

\section{Acknowledgement}
We thank Carlos A.~Salgado for suggesting to look into this topic and Kirill Boguslavski, Tuomas Lappi and Jarkko Peuron for helpful discussions. 
This work has been supported by the Austrian Science Fund FWF No.~P32446-N27 and No.~P28352. 
The Titan\,V GPU used for this research was donated by the NVIDIA Corporation.

\bibliographystyle{elsarticle-num}
\bibliography{references.bib} 

\end{document}